\documentclass[twocolumn]{jpsj3}

\usepackage{amsmath,amssymb,bm}
\usepackage{color}

\title{Mott Transitions in the Hubbard Model with Spatially Modulated Interactions}

\author{Akihisa \surname{Koga}$^1$\thanks{E-mail address: koga@phys.titech.ac.jp},
Takamitsu \surname{Saitou}$^1$, and
Atsushi \surname{Yamamoto}$^2$
}
\inst{$^1$Department of Physics, Tokyo Institute of Technology, 
Meguro, Tokyo 152-8551, Japan\\
$^2$RIKEN, Advanced Institute for Computational Science, 7-1-26 
Minatojima-minami-machi, Chuo-ku, Kobe, Hyogo 650-0047, Japan
}
\abst{
We study two-component fermions in optical lattices 
with spatially alternating on-site interactions
using dynamical mean-field theory.
Calculating the quasi-particle weight, double occupancy, and order parameters 
for each sublattice,
we discuss the low-temperature properties of the system.
When both interactions are repulsive, 
the magnetically ordered state is realized at half-filling. 
In the attractive case,
the superfluid state is, in general, realized with a particle number imbalance. 
On the other hand, when repulsive and attractive interactions are comparable
in the half-filled system,
the magnetically ordered and superfluid states are not realized, but
the metal-insulator transition occurs.
This transition is characterized by the Mott and pairing transitions discussed in
the conventional repulsive and attractive Hubbard models.
In the doped system, 
commensurability emerges owing to the repulsive interactions
and the metal-insulator transition occurs down to quarter-filling.
}

\kword{metal-insulator transitions, spatially modulated interactions}

\begin{document}
\maketitle
\section{Introduction}
Recent extensive experimental and theoretical investigations on ultracold gases 
have been providing a variety of interesting topics. 
Typical examples are ultracold atoms loaded on optical lattices 
with tunable and controllable parameters,\cite{Bloch}
where many unusual phenomena have been observed such as 
the Mott transitions \cite{Greiner,Joerdens,Schneider} and 
the crossover from the BCS-type superfluid to the Bose-Einstein condensation
of tightly bound molecules.\cite{BCSBEC1,BCSBEC2,BCSBEC3}
Recently, the spatial modulation of the atom-atom contact interaction has been 
realized in the bosonic $^{174}$Yb gas system,\cite{Yamazaki} 
and has been stimulating further theoretical investigations 
on particle correlations 
in ultracold atomic systems. 

One of the interesting questions is how the spatial modulation
in the interactions
affects the low-temperature properties of fermionic optical lattice systems.
Here, we consider an optical lattice with 
alternating on-site interactions as the simplest model.
It should capture 
the essence of the systems with modulated interactions
with a finite period.
In our previous study,\cite{proceedings} we clarified that, 
in the half-filled system, single Mott transitions occur when the magnitudes of 
two interactions increase.
However, it is naively expected that the magnetically ordered state or 
superfluid state 
is stabilized in the system only with repulsive or attractive interactions.
Therefore, it is necessary to discuss whether or not the Mott transition is 
indeed realized against magnetic and superfluid fluctuations.
It is also interesting to clarify how alternating interactions 
affect the low-temperature properties of a doped system. 
This should be important in discussing the low-temperature properties 
of realistic optical lattice systems 
where the local particle density varies owing to the existence of 
the harmonic potential.\cite{Joerdens,Schneider,KogaSS,Snoek}
On the other hand, when interacting and noninteracting sites alternate, 
the system is reduced to the two-band model,\cite{Emery,Georges2,ED,Ohashi}
which has been extensively discussed in condensed matter physics.
Thus, our model can be regarded as an extended version of the two-band model.
Therefore, it is instructive to systematically study how low-temperature properties 
are affected by the spatial modulation of on-site interactions.

In this work, we consider the Hubbard model with alternating on-site
interactions to discuss how particle correlations affect low-temperature properties.
To this end, we make use of dynamical mean-field theory (DMFT)
\cite{Metzner,Muller,Pruschke95,Georges96,Kotliar04}
with a continuous-time quantum Monte Carlo (CTQMC) method.\cite{CTQMCRMP}
This combined method is advantageous in dealing with the magnetically ordered and 
superfluid states on an equal footing in the strong-coupling region.
Thus, we will show how such states are affected by alternating interactions.
Furthermore, we discuss the stability of the metal-insulator transition 
in the system.

Our paper is organized as follows.
In \S\ref{2}, we introduce the model Hamiltonian and briefly summarize 
our theoretical approach.
In \S\ref{3}, we demonstrate how the magnetically ordered and 
superfluid states are realized at half-filling.
We clarify that the metal-insulator transition occurs 
in the system with both repulsive and attractive interactions.
We also discuss, in \S\ref{4}, the effect of hole doping to clarify the role of 
repulsive interactions.
We give a brief summary in the last section.

\section{Model and Method}\label{2}

We consider two-component fermions in an optical lattice 
with alternating on-site interactions,
which should be described by the following Hubbard Hamiltonian
on a bipartite lattice:
\begin{eqnarray}
H&=&-t\sum_{\langle ij\rangle\sigma}c_{i \sigma }^{\dagger}c_{j \sigma}
-\sum_{i\sigma} \left(\mu+h\sigma\right)n_{i\sigma}
\nonumber\\
&+&\sum_{i\in A} U_A \left(n_{i\uparrow}-\frac{1}{2}\right)\left(n_{i\downarrow}
-\frac{1}{2}\right)\nonumber\\
&+&\sum_{j\in B} U_B \left(n_{j\uparrow}-\frac{1}{2}\right)\left(n_{j\downarrow}
-\frac{1}{2}\right),
\label{eq:model}
\end{eqnarray}
where $c_{i\sigma}^\dag (c_{i\sigma})$ creates (annihilates) a fermion
at the $i$th site with spin $\sigma$ and 
$n_{i\sigma}=c_{i\sigma}^\dag c_{i\sigma}$. 
$t$ is the hopping integral and
$U_\alpha$ is the on-site interaction at sublattice 
$\alpha\;(=A, B)$. 
$\mu$ is the chemical potential and $h$ is the magnetic field.
Here, we note that both site-dependent interactions and 
potentials have been introduced in the last two terms,
which can be controlled experimentally.\cite{Yamazaki,Peil} 

In our model, there exists a trivial symmetry axis $(U_A=U_B)$, 
which means that the models with $(U_A$, $U_B, \mu, h)$ and 
$(U_B$, $U_A, \mu, h)$ are equivalent to each other. 
When $\mu=h=0\; (\langle n \rangle=1)$, an additional symmetry appears,
where $n=\sum_{i\sigma}n_{i\sigma}/N$
and $N$ is the total number of sites.
By applying the particle-hole transformations~\cite{Shiba}
$c_{i\uparrow}\rightarrow \tilde{c}_{i\uparrow}$ and $c_{i\downarrow}\rightarrow 
(-1)^i \tilde{c}_{i\downarrow}^\dag$
to the model Hamiltonian with $(U_A, U_B, \mu, h)$, we obtain the model Hamiltonian 
with $(-U_A, -U_B, h, \mu)$.
Therefore, in the case with $\mu=h=0$, 
the Hubbard models with $(U_A, U_B)$, $(U_B, U_A)$, $(-U_A, -U_B)$, and 
$(-U_B, -U_A)$ are identical.
In the repulsive model $(U_A>0$ and $U_B>0)$,
a naively expected magnetically ordered state with a staggered moment 
in the $z$-axis ($xy$ plane)
is equivalent to 
the density wave (superfluid) state in the attractive model 
$(U_A<0$ and $U_B<0)$.
It is known that the density wave state is not stable against hole doping.
Therefore, we focus on the stabilities of the magnetically ordered 
and superfluid states in this study.
We note that the particle number imbalance $(n_A\neq n_B)$,
which may be regarded as the density wave state, 
generally appears since the system has 
a two-sublattice structure,
where $n_\alpha=2 \sum_{i\in \alpha,\sigma} 
n_{i\sigma}/N$.
In the following, we focus on the Hubbard model, eq. (\ref{eq:model}),
without the magnetic field, since it is equivalent to the chemical potential
under the particle-hole transformations.

The Hubbard model with alternating interactions has been studied 
in one-dimensional systems, where the possibility of anomalous metallic states 
was discussed.\cite{Kakashvili,Yamamoto}
In this paper,
we deal with the infinite-dimensional Hubbard model to discuss 
how alternating interactions affect low-temperature properties.
To clarify this issue,
we employ the DMFT method.\cite{Metzner,Muller,Pruschke95,Georges96,Kotliar04}
In DMFT, the lattice model is mapped to an effective impurity model, 
where local particle correlations are taken into account precisely.
The lattice Green's function is then obtained via self-consistent conditions 
imposed on the impurity problem. 
The treatment is formally exact in infinite dimensions, 
and even in three dimensions, 
DMFT successfully explains interesting physicical properties such as 
the Mott metal-insulator transition.

When DMFT is applied to the system with a sublattice structure,
Green's function is given 
in the Nambu formalism as\cite{Chitra}
\begin{eqnarray}
{\hat G}({\bf k},z)^{-1}&=&{\hat G}_0({\bf k},z)^{-1}-\hat{\Sigma}(z),
\end{eqnarray}
with
\begin{equation}
{\hat G}_0({\bf k},z)=
\left(
\begin{array}{cc}
 z \hat{\sigma}_0 +\mu\hat{\sigma}_z & -\varepsilon_{\bf k}\hat{\sigma}_z \\
 -\varepsilon_{\bf k}\hat{\sigma}_z&   z\hat{\sigma}_0 +\mu\hat{\sigma}_z 
\end{array}
\right)^{-1},
\end{equation}
and
\begin{equation}
{\hat \Sigma}(z)=
\left(
\begin{array}{cc}
\hat{\Sigma}_A(z) &0 \\
0 & \hat{\Sigma}_B(z) 
\end{array}
\right),
\end{equation}
where $\hat{\sigma}_0$ is the identity matrix, $\hat{\sigma}_z$ is the $z$-component of the Pauli matrix, 
and $\epsilon_{\bf k}$ is the dispersion relation for the bare band.
${\hat \Sigma}_{\alpha}(z)$ is the self-energy for the $\alpha$th sublattice in the Nambu formalism.
The local lattice Green's function is obtained as
$\hat{G}_\alpha(z)=\int dk \hat{G}_{\alpha\alpha}(k,z).$
Here, we use a semicircular density of states,
$\rho(x)=\frac{2}{\pi D}\sqrt{1-\left(\frac{x}{D}\right)^2}$,
where $D$ is the half-bandwidth, which
corresponds to an infinite-coordination Bethe lattice. 
The self-consistency equation is then given by
\begin{eqnarray}
\hat{\cal G}_\alpha (z)&=&z\hat{\sigma}_0 +\mu\hat{\sigma}_z
-\left(\frac{D}{2}\right)^2 \hat{\sigma}_z \hat{G}_{\bar\alpha}(z) \hat{\sigma}_z,
\end{eqnarray}
where $\hat{\cal G}_\alpha (z)$ is the noninteracting Green's function 
of the effective Anderson impurity model for the $\alpha$th sublattice.

There are various numerical methods for solving the effective impurity problem, 
such as the iterative perturbation theory\cite{IPT1,IPT2} and 
exact diagonalization.\cite{ED}
Here, 
to discuss 
low-temperature properties quantitatively,
we use
the CTQMC method.\cite{CTQMCRMP}
This technique has recently been developed \cite{CTQMCRMP} and 
has successfully been applied to general classes of models 
such as the Hubbard model,\cite{Werner,Koga1,Takemori} 
periodic Anderson model,\cite{CTQMCPAM,Luitz}
Kondo lattice model,~\cite{Otsuki} and Holstein-Hubbard model.\cite{Phonon}
Here, we use the hybridization-expansion version of the CTQMC method \cite{Werner} 
extended to the Nambu formalism.\cite{KogaSt}
This allows us to directly access the superfluid state at low temperatures. 
In our CTQMC simulations, we measure normal and anomalous Green's functions 
on a grid of 1000 points.
Furthermore, we also use the numerical renormalization group (NRG) method\cite{NRG}
to discuss the stability of the normal metallic state complementary.
In the NRG, one discretizes the effective bath on a logarithmic mesh 
by introducing the discretization parameter $\Lambda$. 
The resulting discrete system can be
mapped to a semi-infinite chain with exponentially decreasing
couplings, which allows us to access and discuss properties
involving exponentially small energy scales.\cite{KogaNRG}
To ensure that sum rules for dynamical quantities are
fulfilled, we adopt the complete-basis-set algorithm.
\cite{Anders,Peters,Weichselbaum} 
We observe that obeying the sum rules is mandatory
to properly describe the low-energy properties of the
system away from half-filling. In the NRG calculations, we
use the discretization parameter 
$\Lambda=2$ and maintain 4000 states at each step. 

In this work, we use the half-bandwidth $D$ as the unit of energy.
To discuss low-energy properties of the system with alternating interactions 
systematically, we calculate the quasi-particle weight
$z_\alpha=(1-\partial {\rm Re} \Sigma_\alpha(\omega) /\partial \omega)^{-1}$, 
double occupancy $D_\alpha=\langle n_{\alpha\uparrow}n_{\alpha\downarrow}\rangle$, 
magnetization $m_\alpha=(n_{\alpha\uparrow}-n_{\alpha\downarrow})/2$, 
and pair potential 
$\Delta_\alpha=\langle c_{\alpha\uparrow}c_{\alpha\downarrow}\rangle$ 
for each sublattice.
When the CTQMC method is used as an impurity solver,
we calculate the quantity $z_\alpha=(1-{\rm Im} \Sigma_\alpha(i\omega_0) /
\omega_0)^{-1}$ as the quasi-particle weight at finite temperatures,
where $\omega_0=\pi/\beta$.
Furthermore, by applying the maximum entropy method \cite{MEM1,MEM2,MEM3} 
to Green's functions,
we deduce the spectral functions, thus allowing us 
to discuss static and dynamical properties of the systems.

\section{Results at Half-Filling}\label{3}

\begin{figure}[htb]
\begin{center}
\includegraphics[width=7cm]{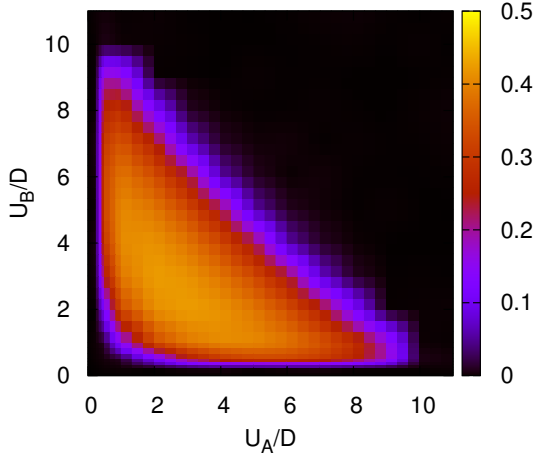}
\caption{(Color online) 
Density plot of the spontaneous magnetization in the system with $T/D=0.05$.}
\label{fig:mag}
\end{center}
\end{figure}

We first discuss the low-temperature properties of the Hubbard model 
at half-filling.\cite{proceedings}
Note that, when $\mu=0$, the local particle density at each sublattice 
is always unity $(n=n_A=n_B=1)$ 
since alternating interactions have been introduced together with 
the alternating potential in the Hamiltonian eq. (\ref{eq:model}).
When the signs of two interactions are the same $(U_A U_B>0)$, 
the ordered ground state is naively expected at zero temperature.
Namely, the magnetically ordered state is stabilized 
in the repulsive case, 
while the superfluid state is stabilized in the attractive case.
Here, we consider the Hubbard model with repulsive interactions,
since the attractive model at half-filling is equivalent 
under the particle-hole transformation, as discussed above.
To clarify how the magnetically ordered state is affected by 
alternating on-site interactions at low temperatures,
we calculate the staggered magnetization $m_{AF}=\sum_i (-1)^i m_i/N$.
The results obtained at $T/D=0.05$ are shown in Fig. \ref{fig:mag}.
In the weak-coupling region, the paramagnetic metallic state is realized 
with $m_{AF}=0$.
Increasing the interactions at both sublattices, 
the staggered magnetization is induced and the phase transition 
occurs to the magnetically ordered state.
In this state, the magnitude of magnetization is almost uniform in the system 
even with alternating interactions $(U_A\neq U_B)$, although the Green's functions 
strongly depend on the sublattices.
A further increase in the interactions leads to another phase transition 
to the paramagnetic state since thermal fluctuations are 
relatively enhanced in comparison with magnetic fluctuations.
Since the effective intersite coupling should be scaled as $t^2/(U_A+U_B)$, 
the phase boundary is almost linear $(U_A+U_B=const.)$ 
at finite temperatures, as shown in Fig. \ref{fig:mag}.

\begin{figure}[htb]
\begin{center}
\includegraphics[width=7cm]{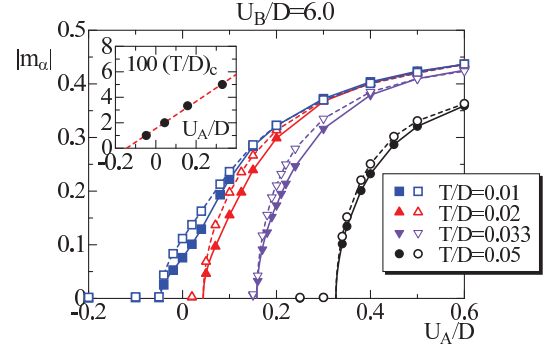}
\caption{(Color online) 
Solid (open) symbols represent the sublattice magnetizations $m_A (-m_B)$ 
in the system with $U_B/D=6.0$
at temperatures $T/D=0.01$ (squares), $0.02$ (triangles), and $0.05$ (circles).
}
\label{fig}
\end{center}
\end{figure}

When the temperature is lowered, the magnetically ordered state becomes more stable
in the weak- and strong-coupling regions.
Here, we discuss the stability of the magnetically ordered state 
when one of the interaction strengths is changed in the weak-coupling region.
In Fig.~\ref{fig}, we show
the spontaneous magnetization in the system 
with a fixed $U_B/D$ of $6.0$ at $T/D=0.01, 0.02$, and $0.05$.
When $U_A/D>0.5$, the repulsive interactions $U_A$ and $U_B$ cooperatively 
stabilize the magnetically ordered state, and 
the sublattice magnetizations are almost the same.
Decreasing the interaction $U_A$, the spontaneous magnetization gradually decreases.
It is found that 
the magnetization for sublattice $B$ is slightly larger than that
close to the critical point.
This implies that the magnetically ordered state is mainly stabilized by 
the larger repulsive interaction $U_B$.
A further decrease in the interaction strength induces the phase transition 
to the paramagnetic state, where the magnetizations simultaneously vanish.
By examining the critical behavior $|m_{AF}| \sim |U_A-(U_A)_c|^\beta$ 
with the exponent $\beta=1/2$, 
we obtain the critical points as
$(U_A/D)_c \sim 0.33, 0.044$, and $-0.047$ at $T/D=0.05, 0.02$, and $0.01$, 
respectively.
By applying the least-squares method to the above critical points,
we determine the quantum phase transition point $(U_A/D)_c$ to be $\sim -0.13$,
as seen in the inset of Fig. \ref{fig}.
This implies that the magnetically ordered state is realized 
even when $U_A U_B<0$,
as long as the attractive interaction is much weaker than the repulsive interaction 
$(|U_A|\ll |U_B|)$ and the temperature is low $(T/D \lesssim 0.01)$.

\begin{figure}[htb]
\begin{center}
\includegraphics[width=7cm]{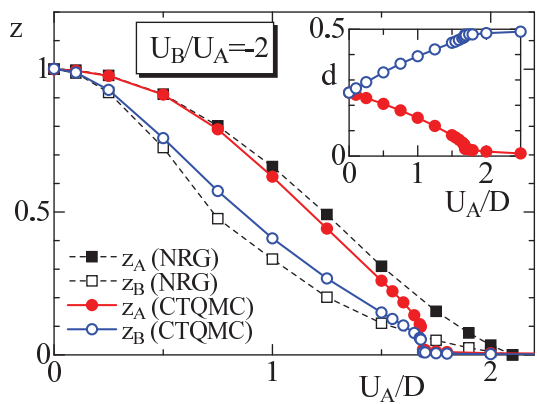}
\caption{(Color online) 
Quasi-particle weight for each sublattice as a function of $U_A/D$ 
with fixed ratio $U_B/U_A=-2$. 
Solid lines with circles are obtained from DMFT by the CTQMC method 
at $T/D=0.02$. 
Dashed lines with squares are obtained from DMFT by the NRG method. 
The inset shows the double occupancy in the system. 
}
\label{fig1}
\end{center}
\end{figure}

On the other hand, when the absolute values of two interactions are comparable, 
a nonmagnetic state is realized.
To discuss the low-temperature properties of the system 
with fixed ratio $U_B/U_A=-2$,
we calculate the quasi-particle weights for both sublattices
at $T/D=0.02$, as shown in Fig. \ref{fig1}.
When the interactions are turned on, the quasi-particle weights $z_A$ and $z_B$
decrease from unity in slightly different ways 
reflecting the difference between on-site interactions.
Namely, a strong renormalization appears in sublattice $B$. 
A further increase in the interactions induces
the jump singularity in both curves at around $U_A/D\sim 1.7$.
Then the double occupancy approaches zero (half)
at sublattice $A\;(B)$, as shown in the inset of Fig. \ref{fig1}.
These imply that the first-order phase transition to the insulating state
occurs, where
the singly occupied states with spin are realized at sublattice $A$,
while the empty or doubly occupied states are equally realized at sublattice $B$.
Therefore, we conclude that the Mott and pairing transitions simultaneously occur 
in sublattices $A$ and $B$, respectively.

\begin{figure}[htb]
\begin{center}
\includegraphics[width=7cm]{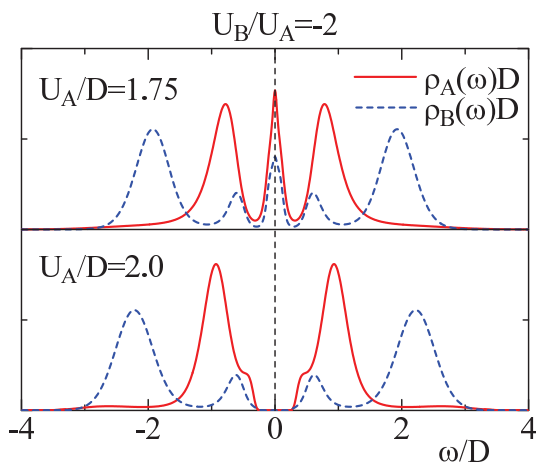}
\caption{
(Color online) Solid (dashed) lines represent the density of states for sublattice $A\;(B)$.
The upper (lower) panel shows the results for the metallic state with $U_A/D=1.75$ 
(the insulating state with $U_A/D=2.0$) in the system 
with $U_B/U_A=-2$ at $T/D=0.02$.
}
\label{rho}
\end{center}
\end{figure}

To confirm this, we show
the density of states at each sublattice at $T/D=0.02$ in Fig. \ref{rho}.
It is found that two kinds of broad peaks appear in the high-energy region.
One of them is located at around $\omega\sim \pm U_\alpha/2$, 
which forms the Hubbard gap due to the local Coulomb interactions.
The other broad peak at around $\omega\sim \pm U_{\bar\alpha}/2$ is 
induced by the on-site interaction at nearest neighbor sites,
although it may be invisible on this scale for the density of states $\rho_A$.
When $U_A/D=1.75$, 
the sharp quasi-particle peaks for both sublattices appear 
near the Fermi level,
which implies that the system is in the metallic state 
close to the Mott transition point.
On the other hand, when $U_A/D=2.0$, 
the gap structure appears in both sublattices, 
which means that the insulating state is induced by 
the Mott and pairing transitions.
Therefore, we conclude that the single metal-insulator transition indeed occurs 
in the system with alternating interactions.

To examine the nature of the phase transition in the paramagnetic state,
we also apply the NRG method as an impurity solver.
In the weak-coupling region, the NRG results are in good agreement with 
the CTQMC results, as shown in Fig. \ref{fig1}.
However, with increasing interactions,
the quasi-particle weights do not have the jump singularity, and
smoothly reach zero at the same critical point $(U_A/D\sim 2.1)$,
which is consistent with the critical point $U_A/D = 3/\sqrt{2}$ obtained
by the linearized DMFT method.\cite{LDMFT} 
This suggests the existence of a second-order quantum phase transition
in the half-filled Hubbard model with alternating interactions,
which is similar to the nature of 
the conventional Mott transition in the paramagnetic state.

We have discussed the low-temperature properties of the infinite-dimensional system 
with attractive and repulsive interactions. 
It has been clarified that 
when both interactions are comparable,
superfluid and magnetic fluctuations are suppressed and 
the paramagnetic phase transition to the insulating state 
with a high residual entropy against any ordered states occurs,
at least, at the temperature $T/D=0.02$.
Since the competition between the superfluid and magnetically ordered states 
can be regarded as a sort of frustration, 
the appearance of the residual entropy in
the insulating state discussed here is similar to that in
the Mott insulating state in the frustrated Hubbard model in infinite dimensions.
By contrast, in the low-dimensional systems, 
the effective next-nearest-neighbor 
interactions between the same sublattices,
which are beyond the DMFT framework based on our local approximation
with two sublattice structures, may realize
the magnetically ordered and/or superfluid (density wave) states.
\cite{Kakashvili}
In this case, its energy scale $t^4/\bar{U}^3$ is much lower than 
the effective on-site interaction $\bar{U}$. 
Therefore, we can say that the metal-insulator transition discussed here is indeed
realized in infinite dimensions, and even in finite dimensions 
it is realized at an intermediate temperature.

In the following, we discuss
how alternating interactions 
affect low-temperature properties in the doped system.

\section{Hole-Doping Effects}\label{4}
In the section, 
we discuss the hole-doping effect on the system with alternating interactions.
First, we consider the system only with repulsive interactions.
It is known that the phase-separated magnetically ordered state 
is realized 
in the conventional Hubbard model 
close to half-filling at zero temperature.\cite{Zitzler2002}
\begin{figure}[htb]
\begin{center}
\includegraphics[width=6cm]{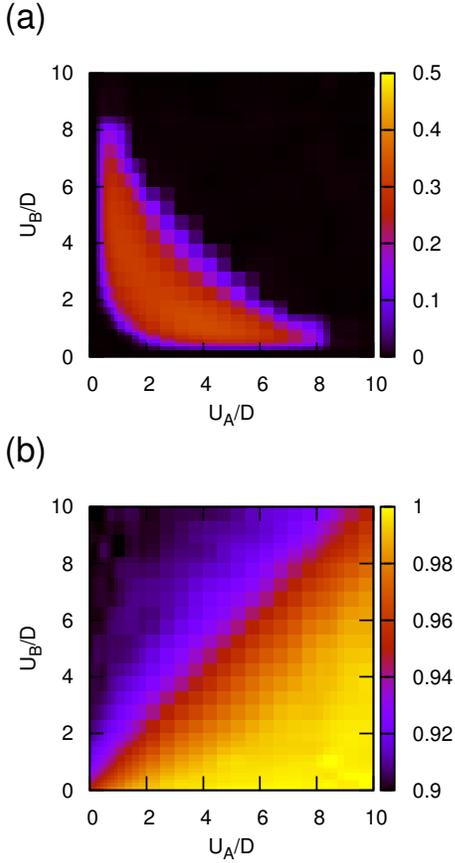}
\caption{(Color online) 
Density plot of spontaneous magnetization (a) and 
particle density in sublattice $A$ (b) for the doped system $(n=0.95)$ 
at $T/D=0.05$.}
\label{fig:mag2}
\end{center}
\end{figure}
To study how the spatial modulation in the interactions affects 
the magnetic structure at low temperatures, 
we calculate the magnetization $m_{AF}$ in the system with $n=0.95$,
as shown in Fig. \ref{fig:mag2}(a).
The magnetization is monotonically decreased by hole doping,
in comparison with the half-filled case (see Fig. \ref{fig:mag}).
It is also found that the phase boundary in the strong-coupling region 
is somewhat curved.
This behavior should be related to the particle occupation in each sublattice.
Figure \ref{fig:mag2}(b) shows the particle density for sublattice $A$.
It is found that commensurability emerges 
in sublattice $A$ $(n_A\sim 1)$ when $U_A\gg U_B$. 
In this region, magnetic correlations are enhanced easily,
which results in the curved phase boundary.
In this case, the DMFT iterations converged well and 
we could not observe phase separation at $T/D=0.05$. 
It is expected that the phase separation occurs
at lower temperatures.\cite{Zitzler2002}
Further hole doping suppresses magnetic correlations,
which leads to the phase transition to the paramagnetic state.

By contrast, it is known that, in the attractive case, 
the superfluid state is stable 
against hole doping and phase separation does not occur.
To clarify how alternating interactions affect 
the superfluid state, we deal with
the attractive Hubbard model with $n=0.75$ and $U_B/U_A=5$.
\begin{figure}[htb]
\begin{center}
\includegraphics[width=7cm]{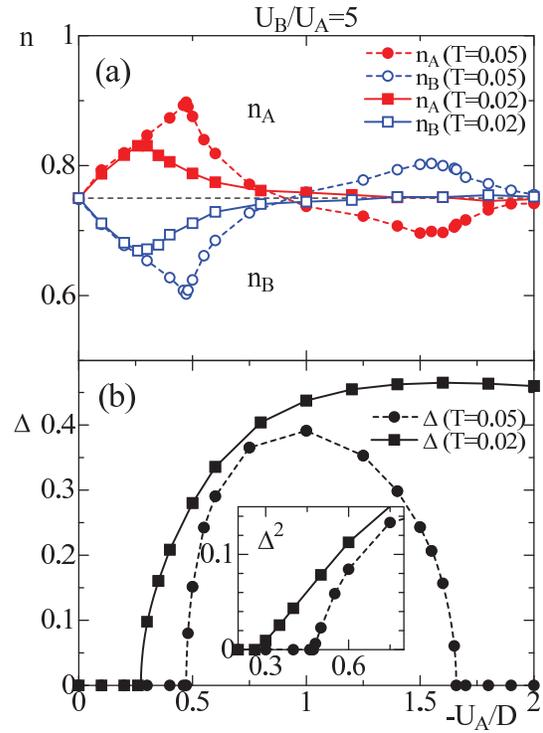}
\caption{(Color online) 
Particle density for each sublattice (a) and pair potential (b) 
as a function of $U_A/D$ with fixed ratio $U_B/U_A=5$. 
Solid (open) symbols are the results for sublattice $A\;(B)$ 
obtained on the basis of DMFT by the CTQMC method. 
}
\label{fig5}
\end{center}
\end{figure}
Calculating the particle density and pair potential,
we obtain the results at fixed temperatures $T/D=0.02$ and $0.05$,
as shown in Fig. \ref{fig5}.
The paramagnetic metallic state is realized below a certain critical interaction 
$(U_A)_c$, 
and the particle density imbalance arises with $n_A\neq n_B$, 
as shown in Fig. \ref{fig5}(a).
In this case, the Hubbard gap $(\sim |U_B|)$ appears  
in the density of states for sublattice $B$,
while a low-energy peak clearly appears in sublattice $A$, 
as shown in Fig. \ref{dos5}.
\begin{figure}[htb]
\begin{center}
\includegraphics[width=7cm]{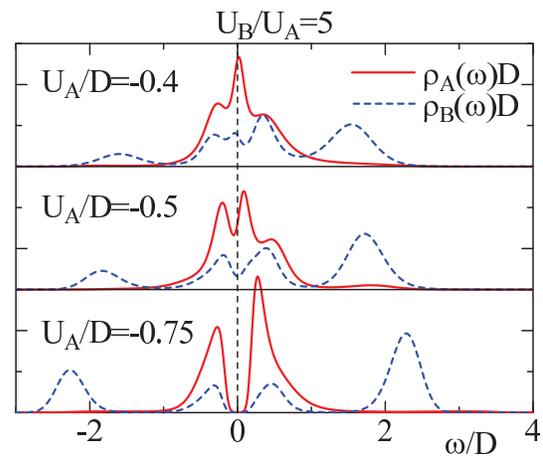}
\caption{(Color online) 
Density of states obtained on the basis of DMFT by the CTQMC method 
for $U_A/D=-0.4, -0.5$, and $-0.75$ at $T/D=0.05$.
}
\label{dos5}
\end{center}
\end{figure}
Beyond the critical interaction $(U_A)_c$, 
the pair potential $\Delta$ is induced and the superfluid state is realized.
Then a dip structure appears in the vicinity of the Fermi level,
as shown in Fig. \ref{dos5}.
Upon decreasing the temperature, the dip structure should continuously change to 
the gap structure.
By examining the critical behavior $\Delta\sim |U_A-(U_A)_c|^\beta$,
we determine the critical interactions
$(U_A/D)_c=0.47$ at $T/D=0.05$ and 
$(U_A/D)_c=0.27$ at $T/D=0.02$,
as shown in the inset of Fig. \ref{fig5}(b).
It is also found that the particle number imbalance is suppressed
once the system enters the superfluid state,
as shown in Fig. \ref{fig5}(a).
Furthermore, the particle number imbalance 
is almost zero when the superfluid order parameter is maximum.
Since this tendency is clearly observed at low temperatures, we can say that
the realization of the superfluid state makes the system uniform.
Similar behavior can be found 
in the half-filled Hubbard model with a uniform attractive interaction
where the hole doping destabilizes the density wave state
and the genuine superfluid state is stabilized.

When $U_AU_B<0$, interesting behavior is exhibited in the doped system ($n=0.75$).
\begin{figure}[htb]
\begin{center}
\includegraphics[width=7cm]{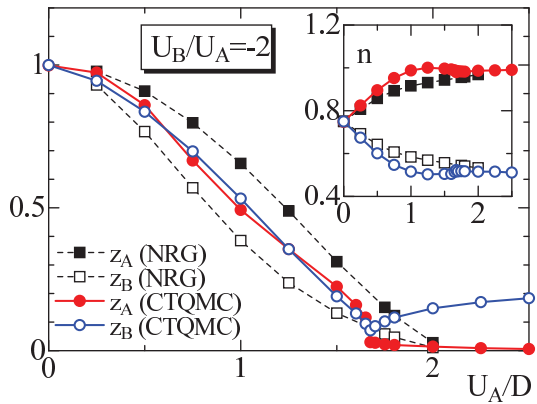}
\caption{(Color online) 
Quasi-particle weight for each sublattice as a function of $U_A/D$ 
with fixed ratio $U_B/U_A=-2$. 
Solid lines with circles represent results obtained on the basis of DMFT by 
the CTQMC method at $T/D=0.02$. 
Dashed lines with squares represent results obtained on the basis of DMFT by 
the NRG method. 
The inset shows the particle density for each sublattice. 
}
\label{fig3}
\end{center}
\end{figure}
The results for the system with the fixed ratio $U_B/U_A=-2$
at $T/D=0.02$ are shown in Fig. \ref{fig3}.
The increase in the interactions monotonically decreases the quasi-particle weight 
for each sublattice and
a singularity appears at $U_A/D\sim 1.7$, 
which is similar to the half-filled case.
However, a further increase in the interaction leads to different behavior.
In sublattice $A$, the quasi-particle weight becomes almost zero and
the local particle density becomes close to half-filling, 
as shown in the inset of Fig. \ref{fig3}.
Our results suggest the existence of the Mott transition in sublattice $A$.
In fact, the Mott gap clearly appears in the density of states, 
as shown in Fig. \ref{dos-d}.
\begin{figure}[htb]
\begin{center}
\includegraphics[width=7cm]{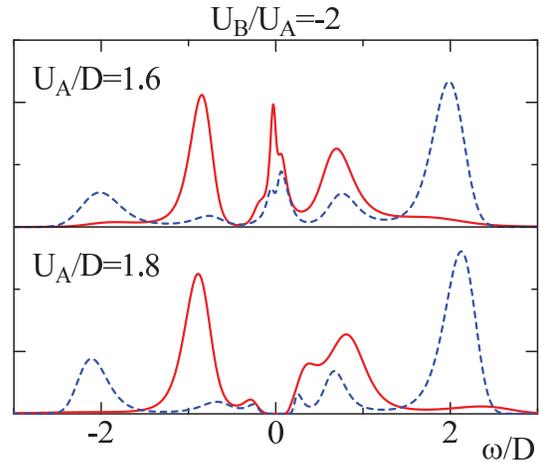}
\caption{(Color online) 
Density of states for the system with $U_B/U_A=-2$ at $T/D=0.02$ 
obtained on the basis of DMFT by the CTQMC method 
when $U_A/D=1.6$ (upper panel) and $1.8$ (lower panel).
}
\label{dos-d}
\end{center}
\end{figure}
By contrast, in sublattice $B$, the local particle density becomes far from 
half-filling. Then the quasi-particle weight  
increases beyond the transition point
and the gap structure appears around the Fermi level.
This behavior is characteristic of the spin-gap state induced by 
the pairing transition in the doped system.\cite{Capone}
Therefore, we can say that even in the doped system,
the Mott and pairing transitions occur at finite temperatures.

However, the nature of the phase transition at zero temperature may not be trivial 
since, in the system with uniform interactions,
the Mott transition is of the second order at half-filling and 
the pairing transition is of the first order far from half-filling.
To clarify the stability of the normal metallic state, 
we calculate the renormalization factors at zero temperature
by the NRG method. 
We see in Fig. \ref{fig3}
that the quasi-particle weights smoothly decrease up to $U_A/D\sim 2.0$.
In this case, a large difference appears between the NRG and CTQMC results,
in contrast to the half-filled case.
This originates from the fact that the particle number at each sublattice 
is sensitive to the interaction strength and thermal fluctuations.
When the interaction is strong enough, oscillation behavior is exhibited
in the DMFT iterations and we do not find a stable solution.
This implies that the first-order phase transition occurs and
the phase separation is realized at zero temperature,
which has been discussed for the attractive Hubbard model.\cite{Capone}
This point thus remains an open issue for future studies.

To clarify the roles of the attractive and repulsive interactions
in the doped system,
\begin{figure}[htb]
\begin{center}
\includegraphics[width=7cm]{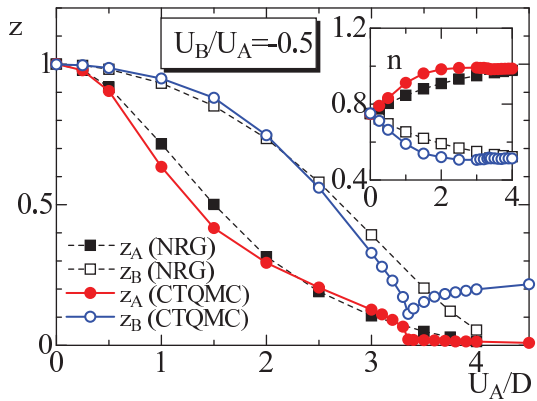}
\caption{(Color online) 
Quasi-particle weight for each sublattice as a function of $U_A/D$ 
with fixed ratio $U_B/U_A=-1/2$. 
Solid lines with circles represent results obtained on the basis of DMFT by 
the CTQMC method at $T/D=0.02$. 
Dashed lines with squares represent results obtained on the basis of DMFT 
by the NRG method. 
The inset shows the particle density for each sublattice. 
}
\label{fig4}
\end{center}
\end{figure}
we also consider the system with $U_B/U_A=-1/2$,
where the magnitude of the attractive interactions is lower than that of
the repulsive interactions.
Figure \ref{fig4} shows that when $U_A/D\sim 1$,
the renormalization factor in sublattice $A$ is rapidly decreased, while
that in sublattice $B$ is hardly decreased.
This behavior is in contrast to that in the case of $U_B/U_A=-2$ 
(see Fig. \ref{fig3}),
and means the absence of the symmetry axis ($U_A=-U_B$).
On the other hand, when the system approaches the phase transition point,
the particle number for sublattice $A$ with repulsive interactions approaches
half-filling
 and both quasi-particle weights decrease. 
Finally, the singularity appears in the curves of the quasi-particle weights
at $U_A/D\sim 3.3$, where
the Mott and pairing transitions occur simultaneously.
Therefore, the nature of the phase transition in the doped system 
is essentially the same as that in the half-filled case.
Note that this is in contrast to the low-temperature behavior
in the doped system with both repulsive interactions.
In such a system, 
the stronger repulsive interaction tends to recover the commensurability
in the corresponding sublattice,
but the weaker one never induces the phase transition 
in the case of intermediate band filling. 
Then the heavy metallic state is realized in the strong-coupling regime.
Therefore, we can say that the single phase transition discussed here 
is stabilized by two effects: 
the Mott transition together with the commensurability 
due to the repulsive interaction
and the pairing transition induced by the attractive interaction.

\begin{figure}[htb]
\begin{center}
\includegraphics[width=7cm]{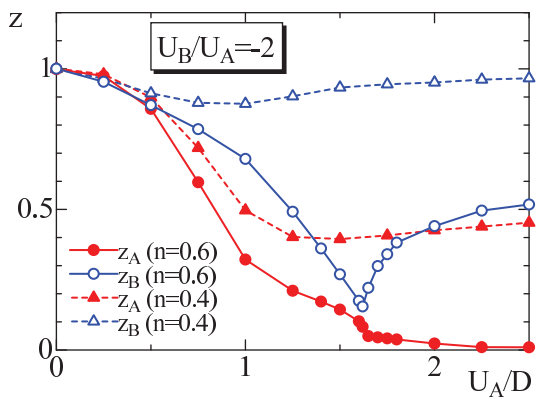}
\caption{(Color online) 
Quasi-particle weight for each sublattice as a function of $U_A/D$ 
with fixed ratio $U_B/U_A=-2$ at $T/D=0.02$. 
Circles (triangles) show the results for $n=0.6\;(0.4)$,
obtained on the basis of DMFT by the CTQMC method.
}
\label{doping}
\end{center}
\end{figure}

When the particle number is smaller than quarter-filling $(0<n<0.5)$, 
it is expected that 
the commensurability never arises and no phase transition occurs.
To confirm this, we show the renormalization factors in the systems
with $n=0.4$ and $0.6$ at $T/D=0.02$ in Fig. \ref{doping}.
In the case of $n=0.6$, a singularity appears at around $U_A/D\sim 1.6$, 
which suggests the existence of the single phase transition.
On the other hand, when $n=0.4$,
no singularity appears in the curves of the renormalization factors. 
Therefore, in the doped system with $n\le 0.5$,
the commensurability never arises and 
no Mott transition occurs.

In this paper, we have discussed how spatially modulated interactions 
affect the low-temperature properties of an optical lattice system,
being motivated by the results of recent experiments 
on the bosonic $^{174}\rm Yb$ gas system.\cite{Yamazaki}
In the case of fermionic optical lattices, on the other hand, 
various parameters are highly controllable,~\cite{Tarruell} and
the spatial modulation in the interaction should 
be an attractive subject of research.
In this sense, our model Hamiltonian eq. (\ref{eq:model}) 
should capture the essence of two-component 
ultracold fermions in optical lattices 
having modulated interactions with a finite period, 
which will be realized in the future.
By studying low-temperature properties systematically, 
we have clarified the role of repulsive and attractive interactions
in the lattice model.
Among them, in the doped system with both interactions, 
we found that the repulsive interaction plays 
a crucial role in stabilizing the metal-insulator transition. 
This is similar to the nature of the orbital-selective Mott transition
in a doped multiband system,\cite{Koga-OSMT,Inaba-OSMT}
where the commensurability in the narrow band arises 
owing to repulsive interactions.

Before closing this section,
we comment on the validity of applying the DMFT method 
to realistic optical lattice systems.
In DMFT, local correlations are correctly treated 
in the effective Anderson impurity model and 
intersite correlations are taken into account at the mean-field level.
Therefore, we consider that
our approach at least qualitatively describes the low-temperature properties of
three-dimensional ultracold fermions, 
except for in the very low temperature region where
the effective next-nearest-neighbor (NNN) interactions become relevant,
as discussed at the end of \S\ref{3}.
An interesting problem is how
the metal-insulator transition discussed in this paper 
competes with the ordered state induced by
the effective NNN interactions in finite dimensions,
which is now under consideration.

\section{Conclusions}
We investigated a fermionic optical lattice system, 
which is described by the Hubbard model with alternating on-site interactions
and potentials.
By combining dynamical mean-field theory 
with the continuous-time Monte Carlo method,
we studied low-temperature properties systematically,
calculating the quasi-particle weight, double occupancy, and order parameters 
for each sublattice.
It was clarified that the magnetically ordered state is realized
in a half-filled system with only repulsive interactions, whereas
the superfluid state is, in general, realized in the attractive case.
We also found that 
when the repulsive and attractive interactions are comparable,
the metal-insulator transitions occur against magnetic and superfluid fluctuations.
In the doped system, 
we found that the commensurability arises owing to the repulsive interactions,
and the insulating state remains at $n>1/2$.

\begin{acknowledgment}
This work was partly supported by the Global Center of Excellence
Program ``Nanoscience and Quantum Physics" of
the Ministry of Education, Culture, Sports, Science and
Technology (MEXT), Japan. 
The simulations were performed using some of the ALPS libraries.
\cite{alps1.3}
\end{acknowledgment}


\end{document}